\documentclass{article}

\usepackage{PRIMEarxiv}

\usepackage[utf8]{inputenc} % allow utf-8 input
\usepackage[T1]{fontenc}    % use 8-bit T1 fonts
\usepackage{hyperref}       % hyperlinks
\usepackage{url}            % simple URL typesetting
\usepackage{booktabs}       % professional-quality tables
\usepackage{amsfonts}       % blackboard math symbols
\usepackage{nicefrac}       % compact symbols for 1/2, etc.
\usepackage{microtype}      % microtypography
\usepackage{lipsum}
\usepackage{fancyhdr}       % header
\usepackage{graphicx}       % graphics
\graphicspath{{media/}}
\usepackage{tabularx}
\usepackage{amsmath}
\usepackage{amsfonts}
\title{An Exploratory Multiome Analysis of a BCG-Associated HSPC Signature: Trained-Immunity Marker in Blood and an Adverse Prognostic Association in Tumour
}

\author{
  Sadia Ruhama \\
  BRAC University \\
\texttt{sadia.ruhama@g.bracu.ac.bd}
}

\begin{document}
\maketitle

\begin{abstract}
Intravesical Bacillus Calmette--Guérin (BCG) is an established treatment for non-muscle-invasive bladder cancer (NMIBC), and recent evidence indicates that bladder BCG induces transcriptional and chromatin changes in hematopoietic stem and progenitor cells (HSPCs). However, the joint multi-omic structure of these BCG-associated changes and its potential to define transferable molecular signatures remain less well characterized. This study performed an exploratory integrative analysis of paired single-cell RNA-seq and ATAC-seq data from two NMIBC patients sampled before and after BCG treatment. Following preprocessing and barcode-level integration, 653 HSPCs were analysed using Multi-Omics Factor Analysis (MOFA+) with 2,000 HSPC-specific variable genes and 25,000 variable chromatin-accessibility peaks. Of ten latent factors, Factor~1 showed the strongest association with BCG status ($r=0.495$) and captured variation across both molecular views. RNA weights from Factor~1 were used to derive candidate gene signatures that showed partial consistency across patient-specific models. Cross-context evaluation showed associations with overall survival in TCGA-BLCA and recurrence in GSE199471, while a subset of signature genes showed consistent directional changes in monocytes following BCG vaccination in GSE184241. Together, these findings identify a candidate multi-omic axis associated with BCG exposure and suggest that its derived molecular signatures exhibit context-dependent behaviour across distinct biological and clinical settings.

\end{abstract}

\section{Introduction}

Intravesical bacillus Calmette--Gu\'erin (BCG) instilled into the bladder is the standard of care for non-muscle-invasive bladder cancer (NMIBC) and reduces recurrence compared with surgical resection alone; however, approximately half of patients have been reported not to maintain a durable response, and no well-established pre-treatment biomarkers of response are available~\cite{daman2025}. How this proportion should be read depends on the clinical endpoint, because recurrence, progression, BCG failure and BCG-unresponsive disease are distinct outcomes, and definitions of BCG-unresponsive disease have been proposed~\cite{fda2018,steinberg2016}. Any statement about BCG response therefore needs to specify which endpoint is meant.

Candidate predictors have been proposed, for example T-cell exhaustion and urinary tumour DNA levels associated with BCG failure~\cite{strandgaard2022}, but the lack of reliable pre-treatment predictors has been attributed in part to an incomplete understanding of how BCG mediates tumour clearance~\cite{daman2025}. As summarised in~\cite{daman2025}, BCG attaches to urothelial cells after instillation, myeloid and lymphoid cells infiltrate the tumour in humans and mice, and in mouse models BCG activates tumour-specific CD4$^{+}$ and CD8$^{+}$ T cells that produce interferon-$\gamma$ and lead to tumour rejection. Tumour-specific CD4$^{+}$ T cells have also been identified in BCG-treated patients~\cite{daman2025}. These effects are presumed to be mediated locally within the bladder, but the upstream events that enable tumour-specific immunity remain poorly defined~\cite{daman2025}.

One candidate upstream mechanism is trained immunity, also termed innate immune memory, in which microbes and their products induce lasting epigenetic changes in innate immune cells that alter their response to later stimuli~\cite{netea2020}. Because many innate immune cells are short-lived, the persistence of these phenotypes has been attributed to changes in hematopoietic stem and progenitor cells (HSPCs) in the bone marrow, which are associated with skewed production of myeloid cells (myelopoiesis) and can be inherited by progeny cells~\cite{daman2025}. In mice, intravenous BCG changes the transcriptional state of hematopoietic stem cells and multipotent progenitors, increases myelopoiesis, and yields macrophages that protect better against \textit{Mycobacterium tuberculosis}~\cite{kaufmann2018}. The effect on progenitors is not specific to BCG, since $\beta$-glucan also modulates myelopoiesis progenitors in mice~\cite{mitroulis2018}.

Evidence in humans comes largely from healthy vaccinated individuals. BCG vaccination of healthy volunteers induced a persistent myeloid-related transcriptional programme in bone-marrow HSPCs, with HNF1 family members implicated~\cite{cirovic2020}. A later study applied single-cell RNA sequencing and sample-paired single-cell ATAC sequencing to bone-marrow CD34$^{+}$ cells from 20 healthy individuals before and after intradermal BCG or placebo. BCG changed both gene expression and chromatin accessibility; expression changes were concentrated in uncommitted stem cells, whereas accessibility changes were concentrated in differentiated progenitors, at sites enriched for KLF/SP and EGR motifs~\cite{sun2024}. Single-cell transcriptomic analysis of blood cells after vaccination has also linked interferon-$\gamma$ signalling to trained-immunity responses~\cite{li2023}. These findings show that BCG-associated changes at the progenitor level can appear in expression and in chromatin accessibility, and that the two layers may change in different cell states.

Observations in cancer patients and tumour models are more limited. In mouse bladder-tumour models, BCG-induced tumour elimination required CD4 and CD8 T cells, and long-term tumour-specific immunity depended mainly on CD4 T cells and tumour-intrinsic interferon-$\gamma$ signalling~\cite{antonelli2020}. Recent studies indicate that bladder administration of BCG can increase circulating cytokines, modify monocyte phenotypes and reduce the frequency of upper respiratory infections, but how these observations relate to anti-tumour immunity, including any role of HSPCs, was unknown~\cite{daman2025}. The trained-immunity studies described above were performed in healthy vaccinated people or in mice, and left open whether bladder-administered BCG reprograms HSPCs in patients with cancer and whether any such reprogramming contributes to anti-tumour immunity~\cite{daman2025}.

This question was addressed directly by Daman et al.~\cite{daman2025}. In a mouse model of bladder cancer, bladder BCG reached the bone marrow, where live BCG was recovered after repeated instillation and confirmed by PCR. It expanded the lineage$^{-}$ Sca-1$^{+}$ Kit$^{+}$ population, reduced the proportion of long-term and short-term hematopoietic stem cells and increased multipotent and common myeloid progenitors, consistent with myeloid-skewed hematopoiesis. HSPCs from BCG-treated mice were sufficient, after transfer to recipients, to confer enhanced tumour control. This depended on interferon-$\gamma$ but not on type I interferon signalling. Myeloid progeny of BCG-experienced HSPCs remodelled the tumour microenvironment, and the resulting tumour control depended on T cells and dendritic cells and was enhanced by PD-1 blockade~\cite{daman2025}.

In patients, Daman et al.\ collected blood from two longitudinal post-resection NMIBC cohorts (8 and 13 patients) before BCG and, as stated in the publication, one week after the fifth instillation~\cite{daman2025}. They profiled circulating CD34$^{+}$ cells by joint single-nucleus RNA and ATAC sequencing of progenitor-enriched blood (PBMC-PIE), a workflow established by the same group~\cite{daman2025,cheong2023}, and obtained 58,652 cells including 1,118 HSPCs.  In HSPCs they reported post-BCG upregulation of antigen-presentation genes and of \textit{BST2}, and predicted transcription-factor activity of AP-1, RUNX and TAL/ZEB/ETS families. In dendritic cells they reported enrichment of interferon regulatory factor (IRF) motif accessibility. Module scores in individual subjects showed upregulation of interferon-$\gamma$ and antigen-presentation programmes in monocytes and dendritic cells, with interindividual variability~\cite{daman2025}. Because HSPCs in this dataset were overrepresented by female cells, the HSPC differential-expression results were dominated by sex-linked genes, and the authors repeated the analysis excluding \textit{XIST}, \textit{TSIX} and Y-chromosome genes~\cite{daman2025}.

The authors are explicit about the limits of their human data. They did not identify how BCG reaches the marrow, they cannot conclude that marrow colonisation is required for the effect, and, given the small human sample, they cannot determine whether measures of innate immune memory predict clinical outcome. They propose as a testable hypothesis that inter-individual differences in pre-treatment or BCG-induced innate immune memory could predict response to BCG~\cite{daman2025}. The paired single-nucleus RNA and ATAC data from these patients were deposited publicly~\cite{daman2025}.

The human analysis in~\cite{daman2025} was modality-specific: it used Scanpy for RNA and ArchR for ATAC, MAST for differential expression and chromVAR for motif accessibility, and a joint latent-factor model is not among the methods listed~\cite{daman2025}. It therefore remains unclear whether a joint decomposition of the paired RNA and ATAC data recovers an axis of BCG-associated variation shared across both layers. This is not a foregone conclusion, because in healthy bone marrow expression and accessibility changes after BCG were reported to be concentrated in different progenitor states~\cite{sun2024}. It is also unclear whether a gene set derived from such an axis behaves consistently in independent datasets. This study did not identify a prior report applying such a model to these data.

Any such analysis is also limited statistically. The discovery data contain few donors, and the original authors themselves note both the small sample and interindividual variability~\cite{daman2025}. Cells from the same donor are not independent replicates, and cell-level tests can overstate confidence when few donors contribute~\cite{squair2021,zimmerman2021}. A cell-level association with BCG status in such data is therefore a candidate for further testing, not an established BCG effect, and technical covariates such as sequencing depth and library quality must be considered alongside it.

Here this study applied MOFA+~\cite{argelaguet2020}, a variational factor model that handles multiple data views and sample groups, to 653 HSPCs from the discovery multiome dataset~\cite{daman2025} (GEO accession GSE295308, the Cohort 2 subseries of GSE295309), sampled before and after bladder BCG. The dataset comprises four paired RNA and ATAC libraries, one per pooled run (R1, R2) per timepoint. RNA expression (2,000 HSPC-specific variable genes) and chromatin accessibility (25,000 variable peaks) were modelled as two views, with the two runs (R1, 295 cells; R2, 358 cells) as groups, and ten factors. We asked whether a single factor is associated with BCG status, and whether that association is separable from technical covariates, particularly ATAC sequencing depth and quality. One factor was associated with BCG status but was also strongly correlated with ATAC fragment count, so we treat it as a candidate axis, not an established BCG programme.

This study then derived versioned gene sets from the positive RNA weights of this factor and compared them with models fitted to each group separately. This study examined the gene sets in four external datasets, among them TCGA-BLCA, a chemotherapy-naive muscle-invasive bladder cancer cohort~\cite{robertson2017}. These datasets differ in tissue, disease stage and BCG exposure, so they bear on how the gene sets behave across contexts, not on BCG response. The study is exploratory and its findings are hypothesis-generating. The remainder of the paper first describes the discovery data, preprocessing, MOFA+ model, gene-set derivation and external datasets; it then reports the factor analysis, the technical-covariate assessment, the gene-set and single-run consistency analyses, and the four external datasets; and it closes with a discussion of how these findings should be interpreted, followed by the limitations, future work and code availability.

% =====================================================================
% Methods, Results, Discussion, Limitations, Future Work
% Items marked \textbf{[CHECK: ...]} could not be resolved from the manuscript or the cited sources.
% Resolve or delete every [CHECK] and [REFERENCE NEEDED] before submission.
% =====================================================================

\section{Methods}

\subsection{Study design}
This exploratory computational study analysed a publicly available paired single-cell RNA and ATAC (10x Multiome) dataset from patients with non-muscle-invasive bladder cancer (NMIBC) sampled before and after intravesical BCG. The workflow comprised preprocessing of each modality, barcode-level integration and identification of hematopoietic stem and progenitor cells (HSPCs), joint latent-factor modelling of HSPCs, assessment of the association of each factor with BCG status and with technical covariates, derivation of gene sets from factor weights, internal consistency analyses, and evaluation of the gene sets in four external datasets. The discovery data comprise two patients. Cells from the same donor are not independent biological replicates~\cite{squair2021,zimmerman2021}, so discovery-stage test statistics computed over cells describe the analysed libraries and are interpreted as hypothesis-generating.

\subsection{Discovery dataset}
 GSE295308 is the primary discovery dataset used in this project, summarises the key properties of this dataset which consists of paired single-cell RNA sequencing and single-cell ATAC sequencing data generated using the 10x Genomics Multiome platform and sequenced on the Illumina NovaSeq 6000. Samples were collected from two NMIBC patients (R1 and R2) at two timepoints, before BCG treatment and after the sixth consecutive bladder BCG administration, yielding eight samples in total (four RNA and four ATAC, one per patient per timepoint). To capture rare circulating hematopoietic stem and progenitor cells, peripheral blood mononuclear cells were enriched for CD34\textsuperscript{+} cells.

\subsection{scRNA-seq preprocessing}
RNA data were processed with Scanpy~\cite{wolf2018}. Cells were retained if they expressed 200--6{,}000 genes and had a mitochondrial read fraction below 20\%. Doublets were identified with Scrublet~\cite{wolock2019}, which flagged 183 cells ($\sim$0.8\%); these were removed. After quality control, 24{,}106 of 24{,}362 cells and 22{,}826 of 36{,}415 genes remained. Data were normalised and 3{,}000 highly variable genes (HVGs) were selected for the full dataset. After HSPC isolation, a separate set of 2{,}000 HSPC-specific HVGs was selected to capture variation within the progenitor population rather than variation dominated by other immune cell types.
. Principal component analysis (50 components) was followed by Harmony~\cite{korsunsky2019} correction of inter-donor technical variation, UMAP embedding and Leiden clustering. The 12 resulting clusters were annotated to 10 immune cell types using canonical markers (\textit{MEIS1} and \textit{FLT3} for HSPCs, \textit{CD14} for classical monocytes, \textit{CD3D} for T cells and \textit{FCER1A} for dendritic cells), and annotations were checked against marker expression, expected cell proportions (HSPCs $\sim$3\%) and the presence of pre- and post-BCG cells in each cluster. 

\subsection{scATAC-seq preprocessing}

Fragment files were imported into SnapATAC2 (v2.9.0)~\cite{zhang2024snapatac2} using the hg38 reference genome, initially retaining barcodes with at least 200 fragments. TSS enrichment scores were calculated in SnapATAC2 using the hg38 gene annotation. Cells were then filtered using the same quality-control criteria in all four libraries: TSS enrichment $\geq 4.0$ and between 1,000 and 50,000 fragments per cell, with the upper fragment threshold used to reduce likely doublets. This retained 55,709 ATAC cells (h1: 30,173; h2: 7,655; h3: 8,454; h4: 9,427). No downsampling was applied.

Peaks were called separately for each library on the retained cells using MACS2~\cite{zhang2008macs}, with ATAC-specific parameters including \texttt{--nomodel}, \texttt{--extsize 200}, \texttt{-p 0.01}, \texttt{--keep-dup all}, and \texttt{--call-summits}. Peaks from the four libraries were concatenated, sorted and merged with bedtools~\cite{quinlan2010bedtools}, and restricted to chromosomes 1--22, X and Y, producing 107,748 consensus peaks from 414,214 raw peaks.

Fragment counts in the consensus peaks were normalised by term frequency--inverse document frequency (TF-IDF)~\cite{cusanovich2015multiplex}. This study selected 50,000 features with SnapATAC2 and computed a 50-component spectral embedding. The first component was excluded to limit the influence of sequencing depth, and components 2--30 were used to construct the 15-nearest-neighbour graph and UMAP embedding.

\subsection{Multimodal integration and HSPC selection}
RNA and ATAC data were matched by cell barcode, which is shared by both assays in 10x Multiome data, and stored in a MuData object~\cite{bredikhin2022}. This yielded 21{,}846 paired cells (90.6\% of RNA and 39.2\% of ATAC barcodes). HSPCs were selected from the RNA-based annotation among these paired cells, giving 653 HSPCs (about 3.0\% of paired cells). Per library, these were 89 (h1, R1 pre-BCG), 206 (h2, R1 post-BCG), 63 (h3, R2 pre-BCG) and 295 (h4, R2 post-BCG), i.e.\ 295 cells in R1 and 358 in R2. Because HSPCs were counted only among cells that passed both RNA and ATAC quality control and could be matched by barcode, this number is smaller than the HSPC count in the source study (1{,}118 HSPCs among 58{,}652 cells), which was obtained from the pooled analysis of two cohorts~\cite{daman2025}. 

\subsection{HSPC-specific features and MOFA+ model}
To reduce dominance of the abundant T-cell and monocyte populations, features were reselected within the HSPC subset: 2{,}000 highly variable genes and 25{,}000 highly variable peaks, both selected with the Scanpy \texttt{seurat\_v3} method (\texttt{span}=0.3) applied to the log-normalised RNA matrix and the TF-IDF-derived ATAC matrix, respectively. The RNA input was not Harmony-corrected; Harmony was used only for embedding and clustering. Both views were modelled with Gaussian likelihoods, without view scaling.

Joint latent-factor analysis used MOFA+ (mofapy2 v0.7.4)~\cite{argelaguet2020}. The model had two views (RNA and ATAC) and two groups (R1, 295 cells; R2, 358 cells). Group-level mean centring (\texttt{center\_groups=true}) was applied, and spike-and-slab priors on the weights (\texttt{spikeslab\_weights=true}) were used to induce sparsity. The model was fitted with 10 factors, at most 2{,}000 iterations, \texttt{convergence\_mode="fast"}, and random seed 42. A single value of $K$ and a single seed were run, and other integration methods were not compared.

\subsection{Factor selection and association with BCG status}
Factor scores were correlated with BCG status (pre-BCG = 0, post-BCG = 1) and with donor (R1 = 0, R2 = 1) by Pearson correlation. A factor was classified as BCG-associated if $|r_{\mathrm{BCG}}|>0.30$ and $|r_{\mathrm{donor}}|<0.5$. The effect-size criterion was added to exclude factors that were significant only because of the number of cells. Alternative BCG-correlation cutoffs were examined for comparison (Table~1).

The selected factor was further summarised by a two-sided Wilcoxon rank-sum (Mann--Whitney) test between post- and pre-BCG cells, with rank-biserial correlation computed as $1-2U/(n_{\mathrm{post}}n_{\mathrm{pre}})$, so that negative values indicate higher scores in post-BCG cells. Cohen's $d$ was computed as the difference in group means divided by the square root of the mean of the two group variances. A logistic regression modelled BCG status (post-BCG = 1) from all ten standardised factor scores jointly (scikit-learn defaults: L2 penalty, $C=1$, \texttt{max\_iter}=1000), fitted to all 653 cells with no held-out set and no donor term. In the permutation test, pre/post labels were shuffled among cells within each donor 1{,}000 times (random seed 42). The test statistic was the Pearson correlation between Factor~1 scores and the shuffled labels, and the two-sided empirical $p$-value was $(k+1)/1{,}001$, where $k$ is the number of permutations with $|r|$ at least as large as the observed value. Variance explained per view and group was obtained from the MOFA+ model.

\begin{table}[h]
\centering
\caption{Factors retained under alternative BCG-correlation cutoffs (HSPC model, 10 factors).}
\begin{tabular}{lll}
\hline
Criterion & Factors retained & Comment \\
\hline
$|r_{\mathrm{BCG}}|>0.30$, $|r_{\mathrm{donor}}|<0.5$ (used) & Factor 1 only & Factor 1: $r=+0.495$ \\
$|r_{\mathrm{BCG}}|>0.50$ & None & Factor 1 ($r=+0.495$) narrowly misses the cutoff \\
\hline
\end{tabular}
\end{table}

\subsection{Assessment of technical covariates}
Each of the 10 factors was correlated (Pearson) with per-cell quality metrics: total ATAC fragments, TSS enrichment, doublet score, total RNA counts and number of detected genes. Mitochondrial fraction was also collected, but its median in HSPCs was 0\% and it had no variance, so it was excluded from the correlation analysis. A factor was flagged if $|r|>0.30$ and $p<0.05$ for any metric. For the BCG-associated factor, an ordinary least squares model was fitted in statsmodels~\cite{seabold2010}:
\begin{equation}
\text{Factor}_1 \sim \text{BCG status} + \log(\text{total fragments}) + \text{TSS enrichment} + \log(\text{RNA counts}) + n_{\text{genes}} .
\end{equation}
Cells were the observations, and the model contained no term for patient or sample. It therefore estimates the association of BCG status with the factor given these covariates, but cannot separate BCG status from sample-level differences.

\subsection{Gene-set derivation}

Gene weights were extracted from the RNA view of the BCG-associated factor (Factor~1). An initial extraction retained the 200 genes with the largest positive weights. A subsequent extraction retained all genes with $|w|>0.01$, yielding 485 genes, of which 275 had positive and 210 had negative weights.

Long non-coding and unannotated transcripts were subsequently removed using a pattern match for accession-style names (\texttt{\^{}[A-Z]\{1,3\}\textbackslash d\{6\}\textbackslash.\textbackslash d+\$}) together with the manually specified genes \textit{LNCAROD}, \textit{FAM30A}, \textit{LINC01122}, and \textit{LINC00926}. This filtering removed 40 genes and left 445 genes, comprising 244 positive- and 201 negative-weight genes.

From the cleaned list, three positive-weight gene sets were defined: a 50-gene core set, a 100-gene standard set, and an extended set containing all 244 positive-weight genes. A signed set contained all 445 genes and was scored as the mean log-normalised expression of the 244 positive-weight genes minus the mean expression of the 201 negative-weight genes. The 50- and 100-gene set sizes were fixed a priori as round-number summaries and were not optimised for performance.

For analyses using the pre-filtered factor-derived list, the 275 positive-weight genes from the 485-gene $|w|>0.01$ set were retained explicitly as a separate set. These 275 genes were used for the rank-window cell-type analysis and for comparison with the patient-specific models and should not be interpreted as the cleaned extended signature.

\subsection{Single-patient models}
A conventional train/validation/test split was not performed, because with two patients cells from one patient do not provide independent test data~\cite{squair2021,zimmerman2021}. Internal consistency was instead examined by refitting MOFA+ separately to the HSPCs of each patient (R1, 295 cells; R2, 358 cells). Each model was a single-group model using the same 2{,}000 variable genes and 25{,}000 variable peaks as the joint model, 10 factors, at most 2{,}000 iterations, spike-and-slab priors on the weights, \texttt{convergence\_mode="fast"} and seed 42. The only difference from the joint model was that group-mean centring was not applied (\texttt{center\_groups=False}). Because the features were selected on all 653 HSPCs, these models are not independent of the joint model.

In each model, the BCG-associated factor was the factor with the largest absolute Pearson correlation with BCG status; no $p$-value or donor criterion was applied. Genes were ranked by signed weight, and the 200 genes with the largest positive weights were taken from each model. Factor sign was not aligned between the two single-patient models. These sets were intersected with the joint-model set, which comprises the 275 positive-weight genes ($|w|>0.01$) of the joint Factor~1, with non-coding genes not removed. The single-patient sets (200 genes each) and the joint set (275 genes) were therefore built by different rules. Overlap was summarised by Jaccard indices, and weights were compared across the 2{,}000 shared genes by Pearson correlation.

\subsection{Cell-type specificity}
The 275 positive-weight genes of the 485-gene file (before removal of non-coding genes) were scored in each annotated cell type using the rank-window score described above (fraction of signature genes among the 1{,}500 highest-expressed genes per cell, computed on the log-normalised matrix). Pre- versus post-BCG scores were compared within each cell type by two-sided Wilcoxon rank-sum tests with rank-biserial correlation $r_b = 1-2U/(n_{\mathrm{post}}n_{\mathrm{pre}})$, so negative values indicate higher scores in post-BCG cells. Cell types with fewer than 10 cells, or fewer than 5 cells in either status, were not tested. $P$-values from the pooled (both-donor) comparisons were corrected across cell types by the Benjamini--Hochberg procedure. The specificity margin was the difference between the absolute $r_b$ in HSPCs and in the next-highest cell type. Effects were also computed separately for each donor; these per-donor $p$-values were not corrected.

\subsection{Motif and pathway analysis}
``BCG-opened'' peaks were the peaks among the 25{,}000 HSPC-variable peaks with a positive Factor~1 ATAC weight above 0.01 (9{,}325 peaks). The background set was drawn from all 107{,}748 consensus peaks excluding the opened peaks, sampled at five times the foreground size within quintile bins of peak width (46{,}786 peaks; seed 42; median width 478~bp versus 481~bp in the foreground). Peaks were not matched for GC content or accessibility.

Motif enrichment was tested with PyJASPAR using the JASPAR 2024 CORE vertebrate collection~\cite{rauluseviciute2024}, scanning peaks with a PSSM bit-score threshold of 8.0, with a one-sided Fisher's exact test and Benjamini--Hochberg correction. 

Over-representation analysis used Enrichr through GSEApy against GO Biological Process 2023 and KEGG 2021 Human, with Enrichr's default background and Benjamini--Hochberg adjusted $p<0.05$. The input was the first 100 genes, ranked by absolute Factor~1 weight, of the 485-gene file after removing accession-style non-coding names; this list contains both positive- and negative-weight genes. Motif enrichment was treated as a hypothesis about regulatory sequence features, not as evidence of transcription-factor activity or binding.

\subsection{External datasets}

Four independent datasets, none of which were used for MOFA+ fitting or gene-set derivation, were examined as cross-context evaluations rather than as conventional held-out validation cohorts.

For TCGA-BLCA ~\cite{tcga_blca_2014}, the 100-gene standard positive-weight signature was evaluated in bulk tumour RNA-seq data. Of the 100 signature genes, 87 were represented in the expression matrix. Expression values were log2$(x+1)$ transformed and the signature was scored per patient using single-sample gene-set enrichment analysis (ssGSEA) implemented in GSEApy~\cite{fang2023gseapy}. Association with overall survival was subsequently examined using Cox proportional-hazards regression and Kaplan--Meier analysis.

GSE176178 ~\cite{Sanders2021BCG} comprised 40 NMIBC patients classified as BCG responders or non-responders. The 50-gene core signature was examined between response groups. Because \textit{XIST} was included in the signature and the response groups were imbalanced by sex, \textit{XIST} was examined separately and sensitivity analyses were performed after removing \textit{XIST} and stratifying the cohort by sex.

For GSE199471 ~\cite{Rouanne2022BCG}, 14 baseline and 11 recurrence samples were analysed using the 200-gene positive-weight signature. Of these genes, 198 were represented in the expression dataset. Expression of each matched gene was standardised across the 25 selected samples, and the standardised values were averaged to obtain one signature score per sample. Baseline and recurrence scores were compared using a two-sided Mann--Whitney U test. Whether baseline and recurrence samples were paired within patients could not be established from the metadata used in the analysis.

GSE184241 ~\cite{Kong2021BCG} was used as a cross-context analysis of circulating monocytes from three healthy donors before and after BCG vaccination. The primary pre-versus-post comparison was restricted to unstimulated (RPMI) monocytes, with LPS-stimulated cells considered separately. Signature-derived gene tiers were compared between pre-vaccination and three-month post-vaccination cells, including donor-specific comparisons. Because statistical tests treated cells as observations despite only three donors, these results were interpreted as descriptive and hypothesis-generating.

\subsection{Statistical analysis}
Tests used were Pearson correlation, Wilcoxon rank-sum and Mann--Whitney tests with rank-biserial correlation, Cohen's $d$, logistic regression, ordinary least squares, permutation testing, Fisher's exact test with Benjamini--Hochberg correction, Cox regression, Kaplan--Meier log-rank tests and AUC. Discovery-stage tests were computed over cells ($n=653$ HSPCs) from two patients and do not account for patient-level clustering. No patient-level aggregated or mixed-effects analysis was performed. Significance was set at $p<0.05$ where a threshold is stated, and no correction across gene-set versions or cohorts was applied.

\section{Results}

\subsection{Discovery cohort and preprocessing}
The discovery dataset comprised four paired RNA and ATAC libraries from two NMIBC patients, one library per patient per timepoint (Table~2). In the RNA modality, 24{,}106 of 24{,}362 cells and 22{,}826 of 36{,}415 genes passed quality control, and Scrublet flagged 183 cells ($\sim$0.8\%) as doublets. Clustering gave 12 clusters, annotated to 10 immune cell types.

In the ATAC modality, 55,709 cells passed quality control after application of the uniform thresholds of TSS enrichment $\geq 4.0$ and 1,000--50,000 fragments per cell. The retained cells comprised 30,173 cells from h1, 7,655 from h2, 8,454 from h3 and 9,427 from h4. Per-sample peak calling produced approximately 75,000--113,000 peaks, and merging 414,214 raw peaks produced 107,748 consensus peaks.

Barcode matching between the independently processed RNA and ATAC modalities yielded 21,846 paired cells. From these paired cells, 653 HSPCs ($\sim$3\%) were identified, comprising 89 R1 pre-BCG, 206 R1 post-BCG, 63 R2 pre-BCG and 295 R2 post-BCG cells. Thus, 295 HSPCs were derived from R1 and 358 from R2.

\begin{table}[htbp]
\centering
\caption{Summary of the GSE295308 discovery dataset used in this study.}
\label{tab:dataset}
\renewcommand{\arraystretch}{1.2}
\begin{tabular}{ll}
\hline
\textbf{Property} & \textbf{Details} \\
\hline
Dataset & GSE295308 (Human Multiome Cohort 2) \\
Study population & 2 NMIBC patients (R1 and R2) \\
Study design & Longitudinal, pre- and post-BCG treatment \\
Cell source & PBMCs enriched for circulating CD34\textsuperscript{+} HSPCs \\
Modalities & scRNA-seq and scATAC-seq (10x Multiome) \\
RNA cells after QC & 24,106 \\
ATAC cells after QC & 55,709 \\
Consensus ATAC peaks & 107,748 \\
Matched RNA--ATAC cells & 21,846 \\
HSPCs analysed & 653 \\
\hline
\end{tabular}
\end{table}

\subsection{Joint factor analysis identifies a candidate BCG-associated factor}

MOFA+ was fitted to the 653 HSPCs with two views (2{,}000 RNA features and 25{,}000 ATAC peaks), two patient groups and 10 factors. The model converged at iteration 36, with the evidence lower bound increasing from $-125{,}380{,}645$ at initialisation to $-410{,}722$; per-iteration values were not recorded. Factor~1 explained 1.77\% and 1.51\% of RNA variance in R1 and R2 and 1.08\% and 1.65\% of ATAC variance. Factor~2 explained more variance than Factor~1, but almost entirely in R1 (6.5\% of RNA and 7.3\% of ATAC variance, against 0.6\% in each view in R2). Factors~3--10 explained at most 0.3\% in any view--group combination. The correlation of the ten MOFA+ factors with per-cell quality metrics in HSPCs is shown in Figure~1.

Among the 10 factors, only Factor~1 met the prespecified association criterion ($|r_{\mathrm{BCG}}|>0.30$ and $|r_{\mathrm{donor}}|<0.5$), with a positive Pearson correlation with BCG status ($r=+0.495$, $p<0.0001$; Table~4). Factor~2 was not associated with BCG status ($r=-0.028$, $p=0.475$). Factors~5 and 6 reached $p<0.05$ with $|r|<0.10$ ($r=-0.093$, $p=0.017$ and $r=-0.095$, $p=0.015$), and the remaining factors had $|r|<0.10$ and $p>0.05$. No factor met a threshold of $|r|\geq0.50$; Factor~1 ($r=0.495$) fell just below it. Factor~1 scores differed between pre- and post-BCG HSPCs (Wilcoxon $p=2.22\times10^{-33}$, rank-biserial correlation $-0.644$; Cohen's $d=1.303$), and its coefficient in a logistic regression on BCG status was $+1.504$, the largest among the 10 factors. In a within-donor permutation test (1{,}000 permutations) the empirical $p$-value was 0.001, the smallest value attainable with this number of permutations (Table~3). These statistics were computed over cells from two patients and do not account for patient-level clustering. Factor~1 is therefore treated as a candidate BCG-associated factor.

The derived signatures were subsequently examined in four independent datasets as cross-context evaluations; these datasets were not treated as conventional held-out validation cohorts (Table~8).

\begin{figure}[h]
\centering
\includegraphics[width=\textwidth]{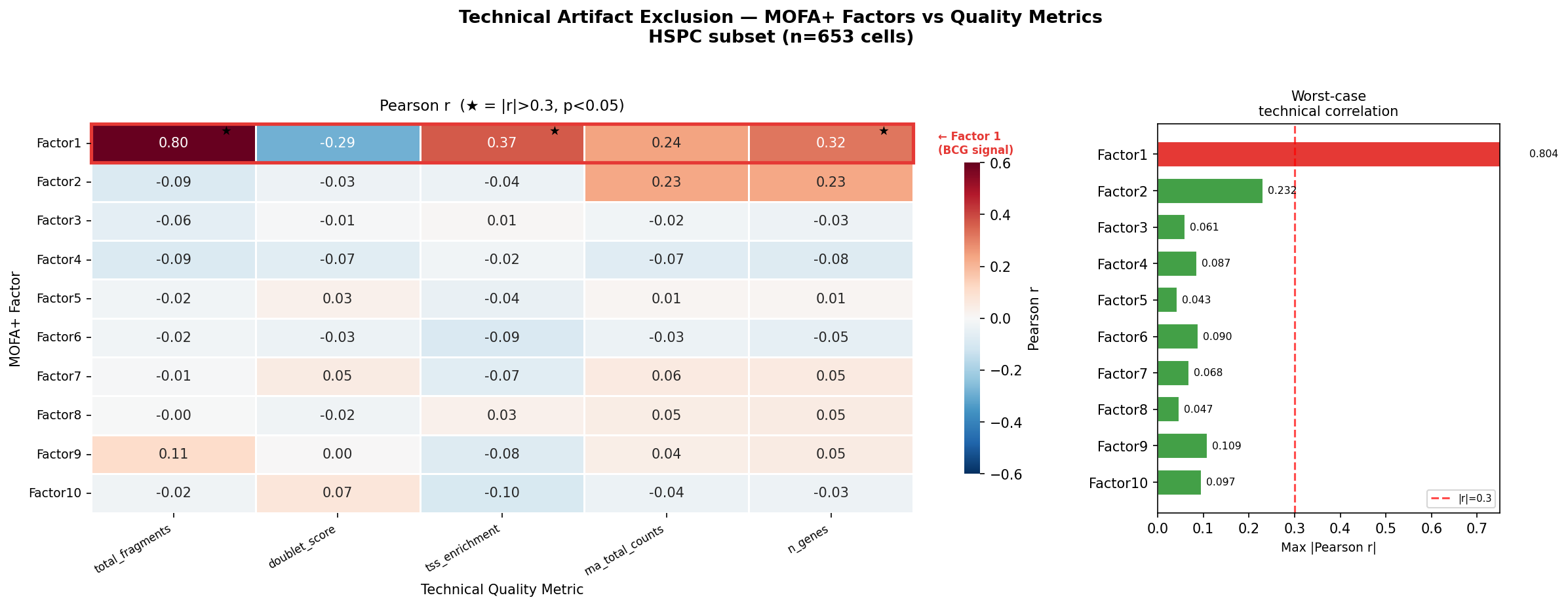}
\caption{Correlation of the ten MOFA+ factors with per-cell quality metrics in HSPCs (n=653). Left: Pearson r; stars mark |r|>0.30 with p<0.05. Right: maximum |r| per factor against any metric. Factor 1 exceeds the threshold for total ATAC fragments (r=+0.80), TSS enrichment (r=+0.37) and number of detected genes (r=+0.32); no other factor exceeds 0.30.}
\label{fig:techheat}
\end{figure}

\begin{table}[htbp]
\centering
\caption{Internal validation metrics for Factor 1 and the derived BCG-associated gene signature.}
\label{tab:internal_validation}
\renewcommand{\arraystretch}{1.2}
\small

\begin{tabularx}{\linewidth}{@{}p{0.25\linewidth} p{0.23\linewidth} X@{}}
\toprule
\textbf{Analysis} & \textbf{Result} & \textbf{Purpose} \\
\midrule

Factor 1 vs BCG status
& $r = +0.495$
& Association with BCG status \\

Multivariable regression
& $\beta_{\mathrm{BCG}} = +0.351$, $R^2 = 0.698$
& Association after adjustment for measured covariates \\

Logistic regression
& AUC $= 0.823$
& In-sample association with BCG status \\

Permutation test
& $p = 0.001$
& Empirical assessment of the BCG--Factor 1 association \\

Rank-window signature score vs Factor 1
& $r = +0.390$, $p = 3.32 \times 10^{-25}$
& Association of the derived gene score with the latent factor \\

\bottomrule
\end{tabularx}
\end{table}

\begin{table}[ht]
\centering
\caption{Association between the ten MOFA+ factors and BCG status.}
\label{tab:factors}
\begin{tabular}{lccc}
\toprule
\textbf{Factor} & \textbf{Correlation ($r$)} & \textbf{$p$-value} &
\textbf{Screening category} \\
\midrule
Factor 1 & $+0.495$ & $<0.0001$ & BCG-associated \\
Factor 2 & $-0.028$ & $0.475$ & No detectable association \\
Factor 5 & $-0.093$ & $0.017$ & Weak association \\
Factor 6 & $-0.095$ & $0.015$ & Weak association \\
Factors 3, 4, 7--10 & $<0.10$ & $>0.05$ & No detectable association \\
\bottomrule
\end{tabular}
\end{table}

\subsection{Technical assessment of Factor 1}
Factor~1 correlated with several technical metrics (Figure~1): total ATAC fragments ($r=+0.804$), TSS enrichment ($r=+0.368$) and number of detected genes ($r=+0.319$). Correlations with total RNA counts (0.24) and doublet score ($-0.29$) were smaller. No other factor had $|r|\geq0.30$ with any metric (maximum 0.232, Factor~2). For Factor~2, which carried variance mainly in R1, the correlation with TSS enrichment was $r=-0.036$ ($p=0.36$).

In an ordinary least squares model of Factor~1 on BCG status, $\log$(total fragments), TSS enrichment, $\log$(RNA counts) and number of genes, BCG status remained associated with Factor~1 (coefficient $+0.351$, $p=2.35\times10^{-23}$). The model explained 69.8\% of the variance of Factor~1 ($R^2=0.698$), and the coefficient of $\log$(total fragments) was $+0.708$. The model treated cells as independent observations and contained no term for patient or sample. BCG status was therefore associated with Factor~1 after adjustment for these covariates, while Factor~1 was also strongly associated with sequencing-depth-related measures; this analysis does not distinguish biological from technical contributions to the depth association.

\subsection{Derivation of the Factor~1 gene signature}

Gene sets were derived from the RNA weights of Factor~1. Applying the $|w|>0.01$ threshold yielded 485 genes (275 positive and 210 negative weights). Removal of 40 long non-coding or unannotated features left 445 genes, comprising 244 positive- and 201 negative-weight genes.

From the cleaned list, three positive-weight signatures were defined: the 50-gene ``BCG up'' core set, the 100-gene standard set, and an extended set containing all 244 positive-weight genes. A signed set containing all 445 genes was scored as the difference between mean expression of the positive- and negative-weight genes.

All signature versions separated pre- from post-BCG HSPCs in the discovery cells. The rank-biserial correlation was $-0.456$ ($p=1.66\times10^{-17}$) for the core set and $-0.450$ ($p=4.06\times10^{-17}$) for the standard set. The signed set showed the largest effect ($r_b=-0.608$, $p=6.23\times10^{-30}$), compared with $r_b=-0.644$ ($p=2.22\times10^{-33}$) for the continuous Factor~1 scores. A rank-window signature score was correlated with Factor~1 ($r=+0.390$, $p=3.32\times10^{-25}$). Because these comparisons used the same cells from which the factor weights were derived, they represent in-sample consistency rather than independent validation.

\begin{table}[htbp]
\centering
\caption{Pre--post BCG associations for Factor~1-derived gene-signature versions in the discovery HSPCs.}
\label{tab:signature_versions}
\renewcommand{\arraystretch}{1.2}
\begin{tabular}{lcc}
\toprule
\textbf{Signature} & \textbf{$p$-value} & \textbf{Rank-biserial $r_b$} \\
\midrule
BCG\_up\_core\_50
& $1.66 \times 10^{-17}$
& $-0.456$ \\

BCG\_up\_standard\_100
& $4.06 \times 10^{-17}$
& $-0.450$ \\

BCG\_up\_top200
& $6.72 \times 10^{-19}$
& $-0.475$ \\

BCG\_signed
& $6.23 \times 10^{-30}$
& $-0.608$ \\

Factor~1 raw score
& $2.22 \times 10^{-33}$
& $-0.644$ \\
\bottomrule
\end{tabular}
\end{table}

\subsection{Cross-patient consistency}

To examine whether the factor depended on one patient, MOFA+ was refitted separately to the HSPCs of each patient. With two patients, this is equivalent to leaving the other patient out and is not cross-validation. The R1-only model (295 cells) converged at iteration 60 and the R2-only model (358 cells) at iteration 57. A BCG-associated factor was identified in each patient-specific model: Factor~1 in R1 ($r=+0.764$) and Factor~1 in R2 ($r=+0.302$) (Table~6).

The intersection of the top 200 genes from each patient-specific model with the 275-gene positive-weight set from the joint model contained 48 genes, including \textit{ETV6}, \textit{ELMO1}, \textit{PLXDC2}, \textit{ATP8B4}, \textit{ANKRD28}, \textit{MED12L}, \textit{ERG}, \textit{ANGPT1}, \textit{SERPINB1}, \textit{INPP4B}, and \textit{PBX1}. The three-way overlap is shown in Figure~2.

Jaccard overlap with the joint-model positive-weight set was 0.146 for R1 and 0.023 for R2 (Table~6). The weight correlation between the two patient-specific models across the 2{,}000 shared genes was $r=-0.330$, and $r=-0.3111$ for signature genes (Figure~3). These results indicate partial but limited cross-patient consistency. Because both patients contributed to the joint model and the features were selected using all 653 HSPCs, this analysis represents an internal consistency assessment rather than independent validation.

\begin{table}[htbp]
\centering
\caption{Patient-specific MOFA+ models and overlap with the 275-gene positive-weight set from the joint model.}
\label{tab:patient_specific}
\renewcommand{\arraystretch}{1.2}
\begin{tabular}{lccc}
\toprule
\textbf{Model} &
\textbf{Selected factor} &
\textbf{BCG correlation} &
\textbf{Jaccard with joint positive-weight set} \\
\midrule

R1
& Factor~1
& $r = +0.764$
& 0.146 \\

R2
& Factor~1
& $r = +0.302$
& 0.023 \\

\bottomrule
\end{tabular}
\end{table}

\begin{figure}[h]
\centering
\includegraphics[width=\textwidth]{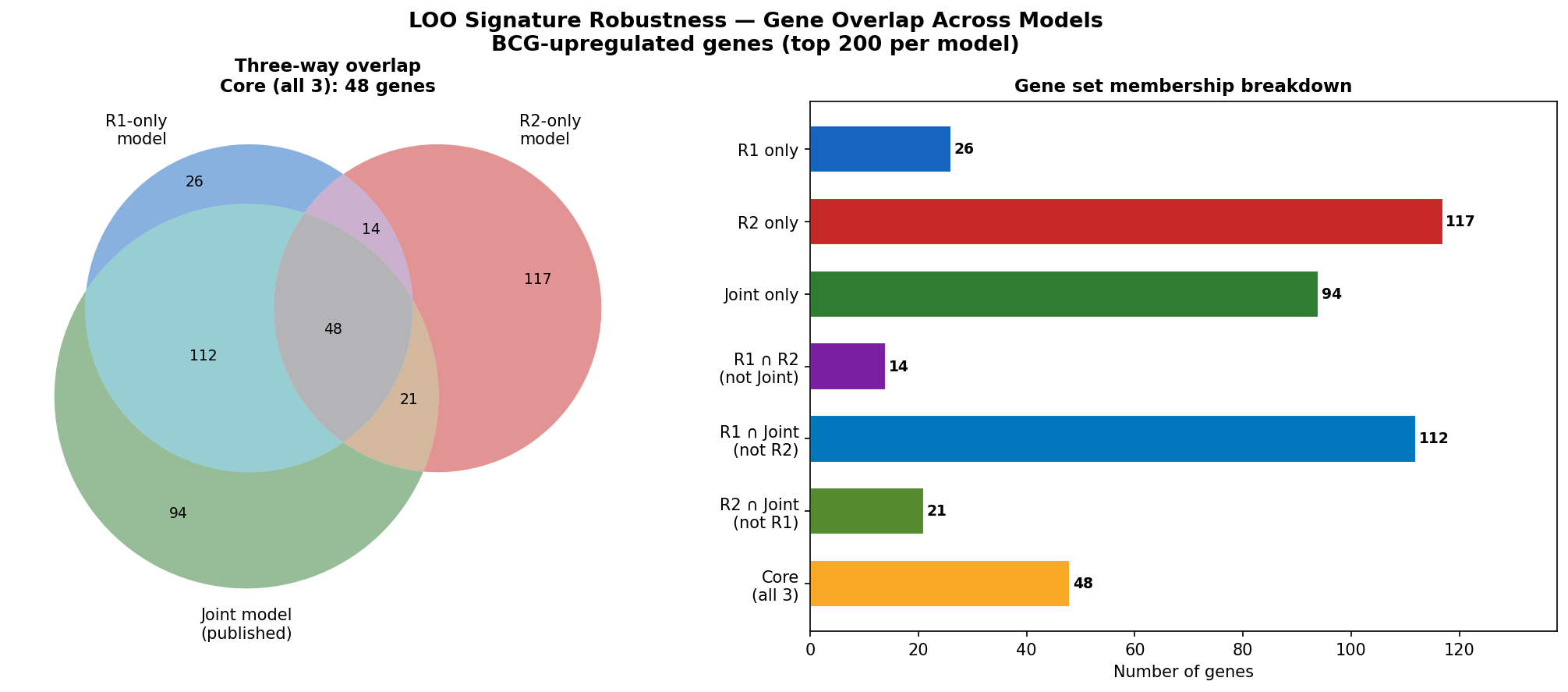}
\caption{Overlap of positive-weight gene sets: R1-only model (200 genes), R2-only model (200 genes) and joint model (275 genes). Counts: 48 genes in all three sets, 112 shared by R1 and joint only, 21 by R2 and joint only, 14 by R1 and R2 only, and 26, 117 and 94 unique to R1, R2 and joint, respectively.}
\label{fig:loo}
\end{figure}

\begin{figure}[htbp]
    \centering
    \includegraphics[width=1.0\textwidth]{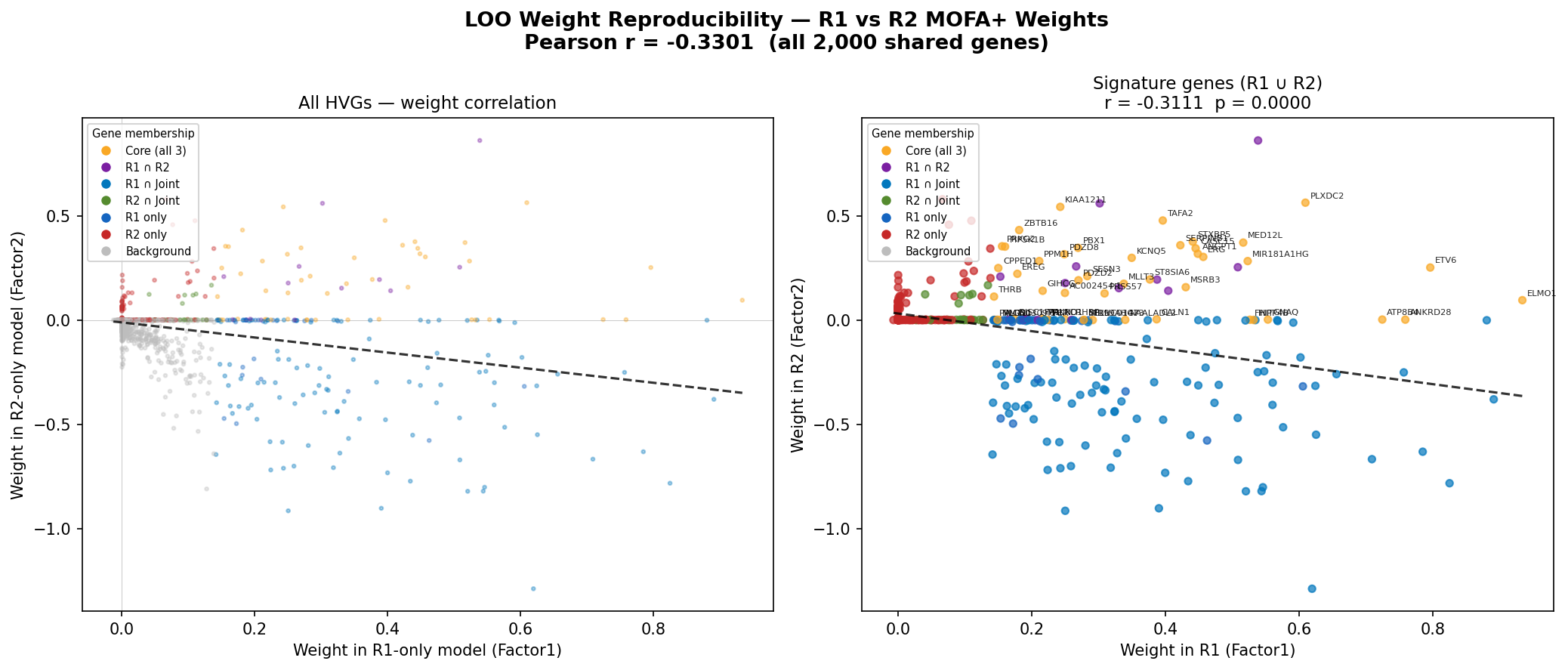}
    \caption{LOO Weight Correlation scatter}
    \label{fig:example}
\end{figure}

\subsection{Cell-type specificity of the Factor~1-derived signature}

To examine the distribution of the Factor~1-derived signature across cell types, a rank-window signature score was calculated for each cell. For each cell, genes were ranked by expression, and the score was defined as the fraction of genes from the 275-gene positive-weight Factor~1 set that occurred among the top 1,500 expressed genes. This score is referred to here as the rank-window signature score and does not represent the AUCell algorithm.

The rank-window signature score was then summarized across annotated cell types to assess the relative distribution of the Factor~1-derived gene set beyond the HSPC population. Differences in score magnitude across cell types were interpreted descriptively and were not used as evidence of HSPC-specificity.

\begin{table}[ht]
\centering
\caption{BCG signature enrichment by cell type.}
\label{tab:specificity}
\begin{tabular}{lll}
\toprule
\textbf{Cell type} & \textbf{$|r_b|$} &
\textbf{\% of HSPC signal} \\
\midrule
HSPC       & $0.521$ & $100\%$ \\
cDC        & $0.355$ & $68\%$  \\
Mono\_CD14 & $0.324$ & $62\%$  \\
Mono\_CD16 & $0.213$ & $41\%$  \\
B\_cell    & $0.171$ & $33\%$  \\
T\_CD4     & $0.090$ & $17\%$  \\
T\_CD8     & $0.059$ & $11\%$  \\
NK         & $0.024$ & $5\%$   \\
\bottomrule
\end{tabular}
\end{table}

\begin{figure}[htbp]
    \centering
    \includegraphics[width=1.0\textwidth]{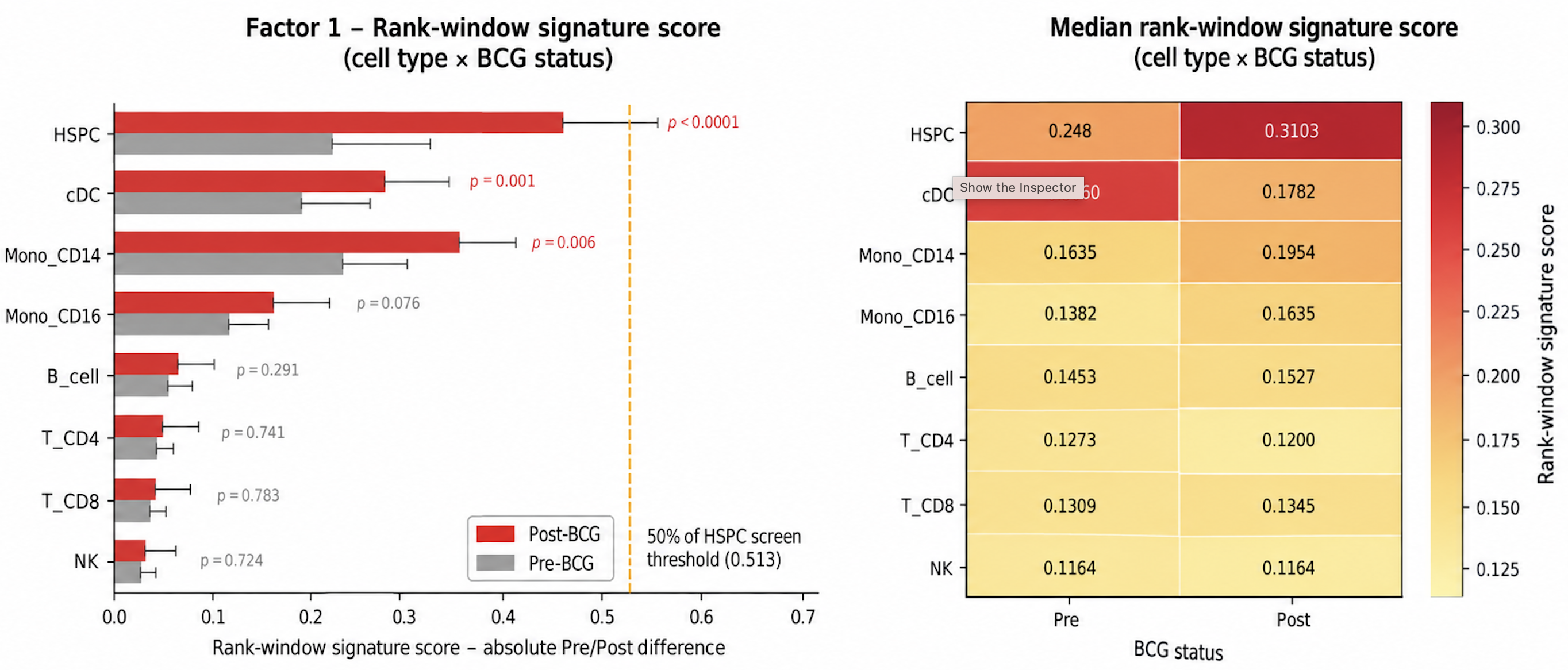}
    \caption{Distribution of the Factor~1-derived rank-window signature score across annotated cell types. Scores represent the fraction of the 275 positive-weight Factor~1 genes occurring among the top 1,500 expressed genes per cell.}
    \label{fig:example}
\end{figure}

\begin{figure}[htbp]
    \centering
    \includegraphics[width=0.8\textwidth]{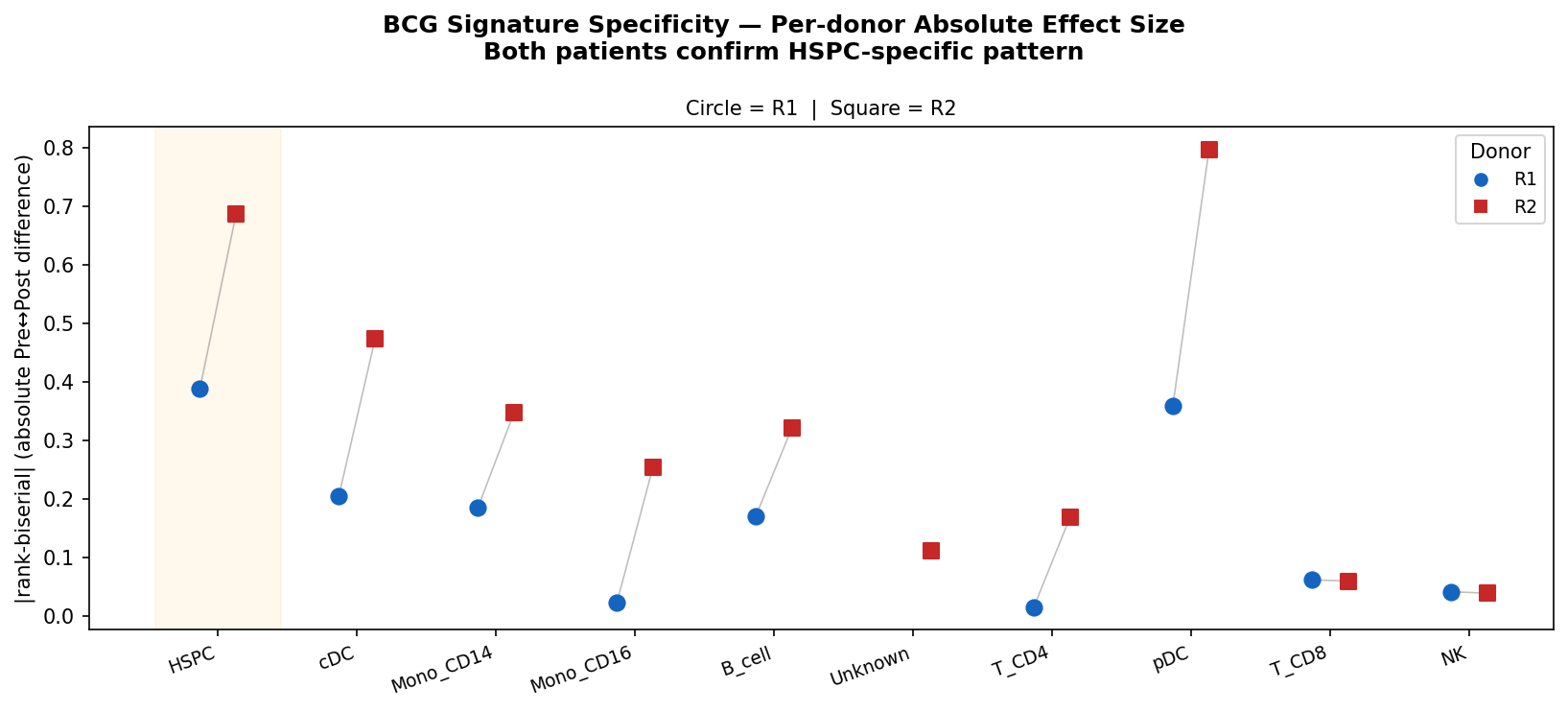}
    \caption{Per-donor absolute effect sizes for pre- versus post-BCG differences in the Factor~1-derived rank-window signature score across annotated cell types. Circles represent R1 and squares represent R2.}
    \label{fig:example}
\end{figure}

\subsection{Motif and pathway analysis}
PyJASPAR analysis of BCG-opened peaks with the JASPAR 2024 CORE vertebrate collection found enrichment of the NF-$\kappa$B/RELB motif (adjusted $p=5.67\times10^{-7}$). Over-representation analysis of the signature genes returned terms related to hematopoiesis, myeloid cell differentiation, lymphocyte differentiation and transcriptional regulation (all adjusted $p<0.05$; Figure~6). Several enriched terms were close to the significance threshold, and some had no apparent connection to hematopoiesis, for example pancreatic and glandular epithelial cell development.

\begin{figure}[htbp]
    \centering
    \includegraphics[width=1.0\textwidth]{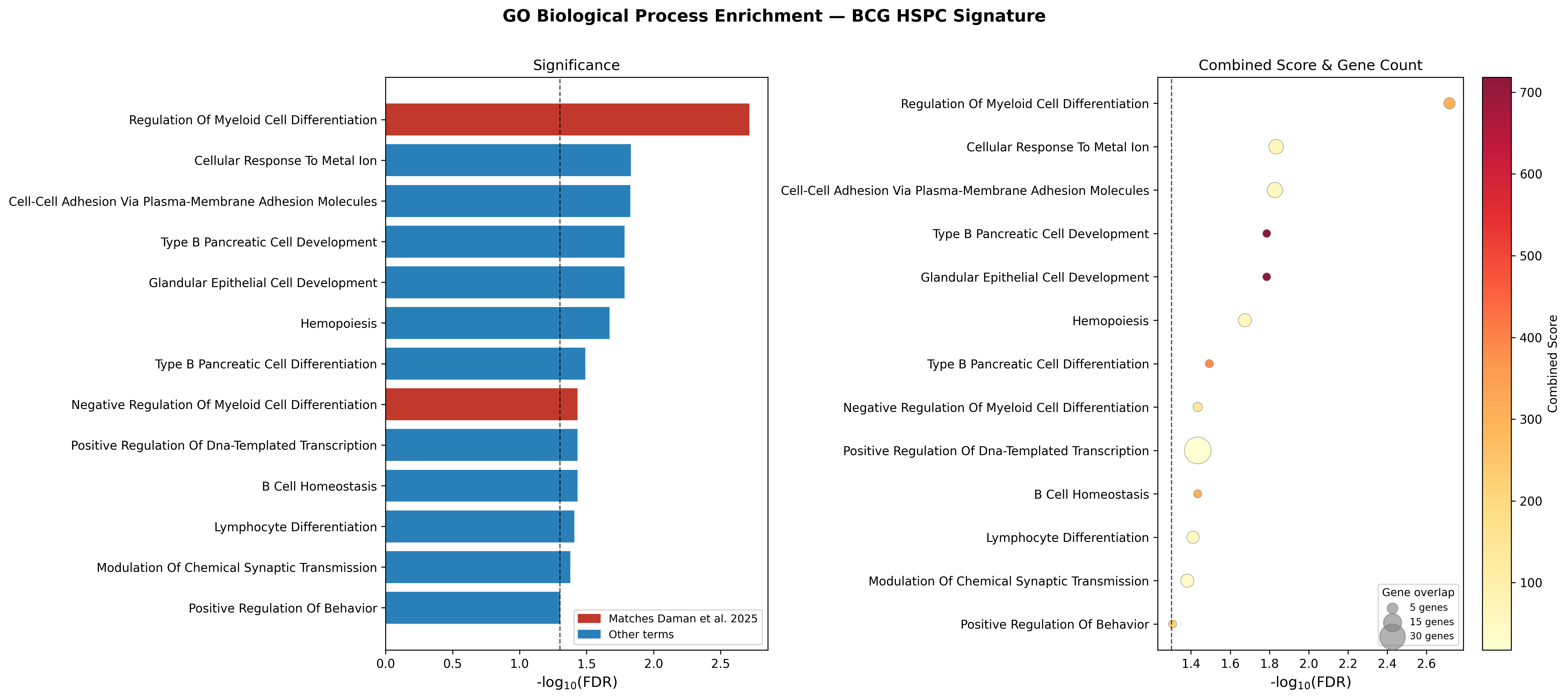}
    \caption{GO BP Enrichment}
    \label{fig:example}
\end{figure}

\subsection{External datasets}
The four external datasets differ in tissue, disease stage and design and are reported separately (Table~8).

\begin{table}[ht]
\centering
\caption{Datasets used for cross-context evaluation of the derived signatures.}
\label{tab:external_cohorts}
\begin{tabular}{p{2.8cm} p{1.5cm} p{3.8cm} p{3.5cm}}
\toprule
\textbf{Cohort} & \textbf{$n$} & \textbf{Design} &
\textbf{Primary metric} \\
\midrule
TCGA-BLCA
    & 407 patients
    & Bulk tumour RNA-seq, OS/PFS
    & Cox HR, KM log-rank $p$ \\[4pt]
GSE176178
    & 40 NMIBC patients
    & BCG responders vs non-responders
    & AUC, Mann-Whitney $p$ \\[4pt]
GSE199471
    & 25 samples
    & Baseline vs recurrence
    & AUC, Mann-Whitney $p$ \\[4pt]
GSE184241
    & 1{,}627 monocytes
    & Healthy donors pre/post BCG vaccine
    & Wilcoxon $p$ per signature tier \\
\bottomrule
\end{tabular}
\end{table}

\subsection{Cross-context evaluation in external datasets}

The four external datasets differed in tissue, disease stage, treatment context, outcome, and study design and were therefore interpreted separately rather than as direct replication cohorts.

\paragraph{TCGA-BLCA.~\cite{tcga_blca_2014}}
The 100-gene standard signature was scored by ssGSEA in 407 patients, with 87 signature genes represented in the expression matrix. Higher signature scores were associated with shorter overall survival (Cox HR $=1.32$, $p=0.046$; Kaplan--Meier log-rank $p=0.014$), with median overall survival of 26.9 months in the high-scoring group and 61.4 months in the low-scoring group. This represents an association between tumour expression of the signature and survival in a distinct bladder-cancer context and does not evaluate BCG response.

\paragraph{GSE176178.~\cite{Sanders2021BCG}}
In 40 NMIBC patients, the 50-gene core signature differed between BCG responders and non-responders ($p=0.043$), with lower scores in responders (mean 1,396 versus 1,790). However, \textit{XIST}, which was included in the signature, showed substantial discrimination by itself (AUC $=0.836$), and seven of the nine female patients were responders. After removal of \textit{XIST} and stratification by sex, the association was substantially attenuated in males and absent in females. This dataset therefore did not provide reliable evidence for a BCG-response association independent of sex, and the initial unadjusted association was considered inconclusive.

\paragraph{GSE199471.~\cite{Rouanne2022BCG}}
Using the initial 200-gene positive-weight signature, 198 genes were represented in the expression dataset. Signature scores were higher in recurrence samples than in baseline samples (mean $+0.172$ versus $-0.135$; two-sided Mann--Whitney $U=30$, $p=0.0109$). The prespecified ROC orientation treated lower scores as indicating recurrence and produced an AUC of 0.195; reversing the orientation to describe the observed direction gives an AUC of 0.805. The latter is descriptive rather than a prespecified estimate of predictive performance. Whether baseline and recurrence samples were paired within patients could not be established from the metadata used in the analysis.

\paragraph{GSE184241.~\cite{Kong2021BCG}}
Of the 100 standard signature genes, 49 were detectable in 1,627 monocytes from three healthy donors before and after BCG vaccination. A ten-gene training tier, including \textit{NRIP1}, \textit{SOX4}, \textit{VIM}, \textit{ELMO1}, and \textit{ATP8B4}, was higher after vaccination ($p=1.04\times10^{-15}$) and showed the same direction in all three donors. \textit{NRIP1}, the highest-weighted Factor~1 gene ($+0.595$), showed an approximately four-fold increase. A 16-gene tier including \textit{HLA-DRA} and \textit{CD74} was lower after vaccination ($p=3.89\times10^{-23}$). Tests treated cells as independent observations despite the presence of only three donors. This analysis therefore provides cross-context directional evidence in circulating monocytes rather than replication of the HSPC analysis.

\subsection{Summary of the evidence}
Findings supporting the candidate signature were the association of Factor~1 with BCG status, the persistence of the BCG-status coefficient after covariate adjustment, partial agreement of the single-patient models, higher scores at recurrence in GSE199471 and the directional consistency of the ten-gene tier in GSE184241. Findings requiring caution were the strong association of Factor~1 with ATAC fragment count, the modest variance explained, the low overlap of the R2-only model with the joint set, the borderline unadjusted TCGA-BLCA result and the dependence of the monocyte analysis on three donors. GSE176178 was inconclusive because of the sex imbalance.

\section{Discussion}

\subsection{Principal findings}
In this exploratory analysis of paired single-cell RNA and ATAC data from HSPCs of two NMIBC patients sampled before and after BCG, one of 10 MOFA+ factors (Factor~1) was positively associated with BCG status ($r=+0.495$) and carried variance in both views (1.08--1.77\% per view per patient). RNA weights from this factor were used to derive versioned candidate gene sets, and 48 genes were shared between the top-ranked sets of the joint model and of models fitted to each patient separately. Results in external datasets were mixed: directionally consistent for a subset of genes in circulating monocytes after BCG vaccination, associated with adverse clinical features in two tumour cohorts, and inconclusive in a third. We therefore interpret Factor~1 and the derived gene sets as a candidate BCG-associated axis and a hypothesis-generating signature, not as an established BCG-induced programme.

\subsection{Interpretation of the multiome factor}
Factor~1 was identified by its association with BCG status, and the group structure and centring of the model meant that it captured variation within patients, not differences between them. BCG status is nevertheless observed at the level of the sample: each patient contributed one pre-BCG and one post-BCG library, and the 653 HSPCs are cells within two donors, not independent biological replicates~\cite{squair2021,zimmerman2021}. The association therefore describes the structure of these libraries and does not establish that BCG caused the factor.

Variance in the ATAC view shows that the factor was not confined to the RNA modality. However, the gene signature was built from RNA weights alone, the contribution of chromatin features to it was not quantified. The variance explained by Factor~1 was modest, and Factor~2 explained more variance (concentrated in R1) without being associated with BCG status. Composition is a further consideration. In mice, bladder BCG shifted hematopoiesis towards myeloid progenitors~\cite{daman2025}, and in healthy human bone marrow BCG-associated expression changes were concentrated in uncommitted stem cells while accessibility changes were concentrated in differentiated progenitors~\cite{sun2024}. HSPCs were analysed here as a single population, so a factor of this kind could partly reflect differences in progenitor-state composition between libraries. This possibility was not tested.

\subsection{Technical and biological interpretation}

Factor~1 correlated more strongly with total ATAC fragment count
($r=+0.804$) than with BCG status ($r=+0.495$), and was also
correlated with TSS enrichment ($r=+0.368$) and the number of
detected genes ($r=+0.319$). In a multivariable model, BCG status
remained associated with Factor~1 after adjustment for the measured
covariates (coefficient $+0.351$; $R^2=0.698$), while
$\log$(total fragments) had a larger coefficient ($+0.708$).

The persistence of the BCG-status term is consistent with variation
in Factor~1 that is not explained by these measured covariates.
However, the model treated cells as independent observations and
contained no patient or sample term. Sample-level or unmeasured
technical effects therefore cannot be excluded.

The strong association between Factor~1 and ATAC fragment count has
several possible explanations that cannot be distinguished by the
present analysis. These include sequencing-depth-related effects,
biological differences in chromatin accessibility, or differences in
progenitor-state composition between samples. No sequencing-depth
matching or RNA-only and ATAC-only sensitivity analyses were performed
to distinguish these possibilities. The association with fragment
count should therefore be considered when interpreting Factor~1 as a
BCG-associated multi-omic axis.

\subsection{Candidate signature and cross-patient consistency}
All signature versions separated pre- from post-BCG HSPCs, with rank-biserial correlations between $-0.450$ and $-0.608$, compared with $-0.644$ for the continuous Factor~1 scores. These comparisons used the cells from which the weights were derived and are in-sample; they show that the gene sets capture part of the factor, not that the signature generalises. The signed set performed best, which suggests that down-weighted genes also carried information, but the data do not indicate why.

Cross-patient consistency was modest and uneven. Overlap with the
joint-model positive-weight set was higher for the R1-only model
(Jaccard $=0.146$) than for the R2-only model ($0.023$), while the
weights of the two patient-specific models were negatively correlated
across the 2{,}000 shared genes ($r=-0.330$). Because the orientation
of a latent factor is not identifiable, a negative weight correlation
need not imply opposite biological effects. The patient-specific models
share data with the joint model, so this analysis is an internal
consistency check rather than independent validation, and with only two
patients it cannot establish population-level reproducibility. The
48-gene intersection is therefore best regarded as a set of candidate
genes.

\subsection{Cell-type and regulatory interpretation}
The pooled pre- versus post-BCG difference in signature scores was largest in HSPCs ($|r_b|=0.521$), followed by cDC (0.355) and CD14 monocytes (0.324). Because the genes were selected from HSPC data, the HSPC effect is not independent evidence of specificity, and in the per-donor analysis the pDC effect was similar to or larger than the HSPC effect, which does not fit a simple HSPC-specific pattern. The data are compatible with an HSPC-centred signal, a broader hematopoietic or progenitor-associated signal, or a mixture, and do not discriminate between them. Signal in myeloid cells would also be compatible with a progenitor-derived programme shared with progeny, as proposed for antigen-presentation programmes in patients~\cite{daman2025}, but this was not tested and the data do not show transmission from HSPCs.

Motif analysis of BCG-opened peaks showed enrichment of the NF-$\kappa$B/RELB motif (adjusted $p=5.67\times10^{-7}$). This indicates that sequence features associated with these factors are over-represented in the selected regions.  Other studies of BCG-associated progenitor changes highlighted different factors: AP-1, RUNX and TAL/ZEB/ETS motifs in HSPCs and IRF motifs in dendritic cells of NMIBC patients~\cite{daman2025}, KLF/SP and EGR motifs in healthy bone marrow~\cite{sun2024}, HNF1 in healthy volunteers~\cite{cirovic2020}, and STAT1 in monocytes after vaccination~\cite{li2023}. The NF-$\kappa$B/RELB result has therefore not been corroborated and is regarded as preliminary. Pathway enrichment (hematopoiesis, myeloid and lymphocyte differentiation) provides functional annotation of the gene set and does not show pathway activation. Interferon-$\gamma$ signalling was prominent in earlier work~\cite{li2023,daman2025} but was not among the enriched terms reported here.

\subsection{Cross-context evidence from external datasets}

The external datasets provided heterogeneous rather than uniform support for the Factor~1-derived signatures, suggesting that their behaviour depends on biological and clinical context. These analyses should therefore be interpreted as cross-context evaluations rather than direct replication of the HSPC discovery analysis.

In TCGA-BLCA, higher scores for the 100-gene standard signature were associated with shorter overall survival (Cox HR $=1.32$, $p=0.046$; log-rank $p=0.014$). Because TCGA-BLCA represents bulk tumour tissue and the outcome analysed was overall survival, this finding indicates an association with clinical outcome in a distinct bladder-cancer context rather than validation of a BCG-response signature.

GSE176178 provided a more cautionary result. Although the 50-gene core signature initially differed between BCG responders and non-responders ($p=0.043$), \textit{XIST} alone showed substantial discrimination and the response groups were imbalanced by sex. Removal of \textit{XIST} and sex-stratified analyses substantially weakened the apparent association. The GSE176178 result was therefore considered inconclusive rather than evidence that the signature distinguishes BCG response.

In GSE199471, the initial 200-gene positive-weight signature was higher in recurrence than baseline samples ($p=0.0109$). The original ROC orientation assumed that lower scores indicated recurrence and consequently yielded an AUC of 0.195; reversing the orientation to reflect the observed direction gives an AUC of 0.805. Because this direction was established after examining the observed scores, the reversed AUC is descriptive and should not be interpreted as a prespecified measure of predictive performance. Nevertheless, the difference in signature scores indicates an association with recurrence status in this dataset.

GSE184241 provided a different form of cross-context evidence. In circulating monocytes from three healthy donors sampled before and after BCG vaccination, a ten-gene training tier derived from the signature increased after vaccination and showed the same direction in all three donors, whereas a 16-gene tier containing \textit{HLA-DRA} and \textit{CD74} decreased. These findings indicate that subsets of the HSPC-derived genes also exhibit BCG-associated directional changes in a distinct immune-cell context. However, the analysis involved only three donors and cell-level statistical testing, and therefore does not constitute independent replication of the HSPC programme.

Taken together, the external analyses indicate context-dependent behaviour of the Factor~1-derived signatures rather than a single uniformly transferable BCG-response pattern. Associations were observed with survival, recurrence, and post-vaccination monocyte states, whereas the direct BCG-response analysis was affected by sex-related confounding. These results support further investigation of the derived signatures across biological contexts while avoiding interpretation as a validated predictive biomarker.

\subsection{Overall implications}
Taken together, the findings are consistent with the possibility that BCG exposure is associated with coordinated expression and chromatin differences in progenitor-associated states, and they motivate testing this hypothesis in larger cohorts. They do not show that BCG induces trained immunity through HSPCs. That would require experiments beyond those performed here, such as the functional work reported in mice~\cite{kaufmann2018,daman2025}. The opposite directions seen in blood-derived cells (higher after BCG) and in tumour cohorts (higher in adverse comparisons) raise a hypothesis about context dependence. The source study proposed that pre-treatment or BCG-induced differences in innate immune memory could predict response~\cite{daman2025}; the present data neither test nor support that hypothesis. The work contributes an exploratory paired RNA and ATAC analysis, a candidate gene set with documented limitations, and a transparent account of how depth-associated latent factors and few-donor designs constrain interpretation.

\section{Limitations and Future work }
\subsection{Limitations}

The discovery analysis included only two NMIBC patients. Although 653 HSPCs were analysed at the cell level, these cells represent observations nested within two donors rather than independent biological replicates. The cell-level statistical analyses therefore do not establish population-level effects, and no patient-level aggregated or mixed-effects analysis was performed. Second, Factor~1 correlated strongly with ATAC fragment count ($r=+0.804$) and moderately with TSS enrichment ($r=+0.368$). Although BCG status remained associated with Factor~1 after adjustment for measured covariates, the regression contained no patient or sample term. Third, the MOFA+ analysis used a single number of factors ($K=10$) and a single random seed, and robustness across alternative model configurations was not evaluated. The threshold used to identify BCG-associated factors was also not established through an independent selection procedure. HSPCs were analysed as a single population without resolving progenitor subtypes, so differences in progenitor-state composition could also contribute to the observed latent structure. Fourth, the external datasets differed substantially from the discovery cohort in tissue, cell type, disease setting, treatment context, and clinical outcome. They therefore represent cross-context evaluations rather than direct replication cohorts.  This study did not identify
an external cohort with comparable longitudinal pre- and post-BCG HSPC
data and matched transcriptomic and chromatin-accessibility measurements.
Consequently, the Factor~1-associated multi-omic pattern could not be
directly tested in an independent, biologically matched cohort. The
available external datasets instead addressed related but distinct
contexts and were therefore used for cross-context evaluation rather
than direct replication.

Technical confounding remains a limitation because Factor~1 was associated with ATAC fragment count and TSS enrichment, and no patient-level or sequencing-depth-matched sensitivity analysis was performed. The derived signatures were evaluated internally in the discovery data, and no suitable independent longitudinal HSPC dataset was identified for direct replication. The available external datasets differed in biological and clinical context and therefore provide cross-context rather than direct validation; notably, GSE176178 was affected by sex/\textit{XIST} confounding, GSE199471 had unclear sample pairing, and GSE184241 included only three donors.

\subsection{Future Work}

Future studies should evaluate the Factor~1-associated programme in larger longitudinal NMIBC cohorts containing sufficient numbers of independent patients and matched pre- and post-BCG samples. Donor-aware statistical approaches, including patient-level aggregation or mixed-effects modelling, will be important for distinguishing reproducible treatment-associated changes from cell-level variation.  Matching or otherwise controlling ATAC sequencing depth, varying the number of MOFA+ factors and random seeds, and resolving HSPC subtypes could further assess the robustness of Factor~1 and distinguish technical variation from differences in progenitor-state composition. The derived gene signatures should also be evaluated prospectively in independent cohorts designed specifically to assess BCG treatment and clinical response, with relevant covariates incorporated into the analysis. Cross-context findings from tumour and circulating immune-cell datasets can be used to generate hypotheses, but direct validation in appropriately matched NMIBC cohorts will be necessary to determine the generalisability and potential clinical relevance of the signatures.

\section{Conclusion}

This study applied joint latent-factor modelling to paired single-cell RNA-seq and ATAC-seq data from HSPCs of NMIBC patients sampled before and after intravesical BCG treatment. MOFA+ identified Factor~1 as the latent factor most strongly associated with BCG status, capturing coordinated variation across the transcriptional and chromatin-accessibility views. RNA weights from this factor yielded candidate gene signatures that showed partial consistency across patient-specific models and the strongest pre--post BCG difference in HSPCs relative to the other analysed cell types. Evaluation in independent datasets revealed context-dependent behaviour of the derived signatures. Higher signature scores were associated with shorter overall survival in TCGA-BLCA and with recurrence in GSE199471, while a subset of signature-derived genes showed consistent directional changes in circulating monocytes following BCG vaccination in GSE184241. In contrast, the apparent BCG-response association in GSE176178 was substantially affected by sex-related confounding and was considered inconclusive. Together, these findings identify a candidate multi-omic axis associated with BCG exposure and provide a framework for deriving molecular signatures from coordinated transcriptional and chromatin variation. Given the small discovery cohort and the strong association of Factor~1 with ATAC fragment count, the identified axis and its derived signatures should be regarded as hypothesis-generating. Validation in larger, donor-aware and independently collected NMIBC cohorts will be necessary to determine their biological robustness and potential clinical relevance.

\textbf{Code availability}
The analysis code for reproduction is available at \url{https://github.com/sadiaruhama/MultiomeLatent}.

\textbf{Data availability}
All datasets analyzed in this study are publicly available from the corresponding public repositories and accession numbers cited in the manuscript.

%Bibliography
\bibliographystyle{unsrt}  
\bibliography{references}

\end{document}